\documentclass[conference]{IEEEtran}
\IEEEoverridecommandlockouts
\usepackage{cite}
\usepackage{amsmath,amssymb,amsfonts}
\usepackage{algorithmic}
\usepackage{graphicx}
\usepackage{textcomp}
\usepackage{xcolor}
\usepackage{algorithm}
\usepackage{algorithmic}

\usepackage{booktabs}    
\usepackage{multirow}    
\usepackage{array}       

\def\BibTeX{{\rm B\kern-.05em{\sc i\kern-.025em b}\kern-.08em
    T\kern-.1667em\lower.7ex\hbox{E}\kern-.125emX}}
\begin{document}

\title{Sampling Strategy Design for Model Predictive Path Integral Control on Legged Robot Locomotion\\
}



\author{
\IEEEauthorblockN{1\textsuperscript{st} Chuyuan Tao}
\IEEEauthorblockA{\textit{School of Artificial Intelligence} \\
\textit{Shandong University} \\
Jinan, China \\
202599000126@sdu.edu.cn}
~\\
\and
\IEEEauthorblockN{2\textsuperscript{nd} Fanxin Wang*}
\IEEEauthorblockA{\textit{Department of Mechatronics and Robotics} \\
\textit{Xi'an Jiaotong-Liverpool University} \\
Suzhou, China \\
Fanxin.Wang@xjtlu.edu.cn}
*Corresponding author
~\\
\and
\IEEEauthorblockN{3\textsuperscript{rd} Haolong Jiang}
\IEEEauthorblockA{\textit{Department of Mechatronics and Robotics} \\
\textit{Xi'an Jiaotong-Liverpool University} \\
Suzhou, China \\
\textit{University of Liverpool} \\
United Kingdom\\
Haolong.Jiang22@student.xjtlu.edu.cn}
~\\
\and
\IEEEauthorblockN{4\textsuperscript{th} Jia He}
\IEEEauthorblockA{\textit{Department of Mechatronics and Robotics} \\
\textit{Xi'an Jiaotong-Liverpool University} \\
Suzhou, China \\
\textit{University of Liverpool} \\
United Kingdom\\
Jia.He20@student.xjtlu.edu.cn}
~\\
\and
\IEEEauthorblockN{5\textsuperscript{th} Yiyang Chen}
\IEEEauthorblockA{\textit{Mechatronics and Robotic Systems} \\
\textit{Xi'an Jiaotong-Liverpool University} \\
Suzhou, China \\
Yiyang.Chen2302@student.xjtlu.edu.cn}
~\\
\and
\IEEEauthorblockN{6\textsuperscript{nd} Qinglei Bu}
\IEEEauthorblockA{\textit{Department of Mechatronics and Robotics} \\
\textit{Xi'an Jiaotong-Liverpool University} \\
Suzhou, China \\
Qinglei.Bu02@xjtlu.edu.cn}
}

\maketitle

\begin{abstract}
Model Predictive Path Integral (MPPI) control has emerged as a powerful sampling-based optimal control method for complex, nonlinear, and high-dimensional systems. However, directly applying MPPI to legged robotic systems presents several challenges. This paper systematically investigates the role of sampling strategy design within the MPPI framework for legged robot locomotion. Based upon the idea of structured control parameterization, we explore and compare multiple sampling strategies within the framework, including both unstructured and spline-based approaches. Through extensive simulations on a quadruped robot platform, we evaluate how different sampling strategies affect control smoothness, task performance, robustness, and sample efficiency. The results provide new insights into the practical implications of sampling design for deploying MPPI on complex legged systems.
\end{abstract}

\section{Introduction}

MPPI control has recently become a highly effective sampling-based optimal control framework for handling complex, nonlinear, and high-dimensional systems~\cite{williams2017information,williams2018information,theodorou2010generalized}. By leveraging Monte Carlo rollouts and importance sampling, MPPI avoids explicit gradient computation and naturally accommodates nonlinear dynamics, nonconvex cost functions, and hard-to-model interactions. A key advantage of MPPI is its inherent compatibility with parallel computing architectures, such as GPUs, which enable thousands of trajectory rollouts to be evaluated simultaneously~\cite{williams2015model,lowrey2018plan}. This property makes MPPI particularly attractive for systems with complex dynamics and hybrid behaviors, including legged robots, where analytical solutions and convex formulations are often impractical~\cite{hutter2016anymal,rathod2021terrain}.

Despite these advantages, directly applying MPPI to legged robotic systems presents several challenges. One major limitation lies in the control sampling process itself. Standard MPPI implementations typically sample control inputs independently at each time step, which can lead to temporally uncorrelated and non-smooth control sequences~\cite{williams2016aggressive}. For legged robots operating in contact-rich environments, such non-smooth control signals may result in unstable behaviors, excessive actuator effort, or poor contact consistency. These issues are further exacerbated by the hybrid nature of legged locomotion dynamics, where discontinuities due to contact switching and impact events pose significant challenges for control and optimization~\cite{chatzinikolaidis2020contact,gangapurwala2022rloc}.

To address these difficulties, prior work has explored structured control parameterizations within sampling-based and MPC frameworks. Trajectory representations based on splines, basis functions, or reduced-order parameterizations have been shown to improve smoothness and sample efficiency in optimal control~\cite{betts2010practical,tassa2014control}. Notably, the work \emph{``Real-Time Whole-Body Control of Legged Robots with Model-Predictive Path Integral Control''}~\cite{alvarez2025real} demonstrated a successful real-time MPPI implementation on legged robots by sampling control trajectories in a lower-dimensional space using cubic spline parameterization. Instead of directly sampling discrete control actions at each time step, the authors parameterized control sequences with spline coefficients, thereby enforcing temporal smoothness and reducing the effective dimensionality of the sampling space. This approach significantly improved control smoothness and enabled stable whole-body behaviors in real-world experiments.

Inspired by this line of work, this paper systematically investigates the role of sampling strategy design in MPPI for legged robot control. Building upon the idea of structured control parameterization, we explore and compare multiple sampling strategies within the MPPI framework, including both unstructured and spline-based approaches. Through extensive simulations on a legged robot platform, we evaluate how different sampling strategies affect control smoothness, task performance, robustness, and sample efficiency. The results provide new insights into the practical implications of sampling design for MPPI-based legged locomotion and highlight sampling strategy as a critical factor in deploying MPPI on complex robotic systems.

\subsection{Related Work}

Sampling-based optimal control methods have been widely studied for robotic systems operating under nonlinear dynamics, uncertainty, and safety constraints~\cite{theodorou2010generalized,todorov2009efficient}. Among these approaches, path-integral-based control methods, such as Model Predictive Path Integral (MPPI), have attracted significant attention due to their ability to handle nonconvex optimal control problems through stochastic sampling and parallel computation~\cite{williams2017information,testouri2023towards}. Recent research has focused on improving the safety and robustness of path-integral control by incorporating control-theoretic structures into the sampling process. Tao \emph{et al.}~\cite{tao2022path} proposed a framework that integrates stochastic control barrier functions (CBFs) into path-integral control, enabling probabilistic safety guarantees while preserving computational efficiency. Similarly, Wang \emph{et al.}~\cite{wang2025dbas} introduced DBaS-Log-MPPI, which augments MPPI with barrier states to shape the sampling distribution and improve safety and efficiency in constrained environments.

In parallel, informed sampling has also been shown to be essential in motion planning and learning-based control for complex environments. Tao \emph{et al.}~\cite{tao2025guided} developed a guided sampling-based motion planning algorithm for dynamic and obstacle-cluttered environments, highlighting how structure in the sampling process can significantly improve planning efficiency and robustness. Related ideas have emerged in reinforcement learning, where safety and robustness are enforced through adaptive control and barrier-based mechanisms~\cite{cheng2022improving,cheng2023safe}. Although learning-based approaches typically rely on offline training, they share a common insight with sampling-based optimal control: imposing structure on the action or control space is crucial for reliable performance. For legged robots in particular, prior work has shown that directly sampling time-discretized control inputs often leads to nonsmooth and unstable behaviors, motivating the use of structured parameterizations such as splines or reduced-order representations in MPC and MPPI implementations~\cite{alvarez2025real,rathod2021terrain}. In contrast to introducing a single structured representation, this work systematically compares multiple sampling strategies within the MPPI framework to better understand their impact on performance and robustness in legged locomotion tasks.

\section{Preliminaries}
\label{sec:preliminaries}

\subsection{Experimental Platform and Tasks}

All experiments in this work are conducted on a Unitree quadruped robot equipped with torque-controlled joints and onboard state estimation. The robot provides
full-state feedback including base pose, velocity, and joint states, enabling
real-time whole-body control under dynamic locomotion tasks.

We evaluate sampling strategies within the Model Predictive Path Integral (MPPI)
framework on a set of representative locomotion scenarios that stress different
aspects of control smoothness, adaptability, and robustness:
\begin{itemize}
\item \textbf{Flat-ground walking}: steady-state locomotion on level terrain,
used to evaluate nominal tracking performance and control smoothness.
\item \textbf{Stair climbing}: traversal of a staircase with discrete height
changes, requiring precise foot placement and coordinated whole-body motion.
\item \textbf{Large obstacle (box) traversal}: stepping over a large box obstacle,
which induces abrupt changes in contact configuration and demands expressive
control trajectories.
\end{itemize}

These tasks are commonly used benchmarks for legged locomotion and jointly
capture steady, moderately dynamic, and highly non-smooth interaction regimes.
All sampling strategies are evaluated under identical task settings, cost
functions, and control constraints to enable a fair comparison.

\subsection{Model Predictive Path Integral Control}

Model Predictive Path Integral (MPPI) control is a sampling-based stochastic
optimal control method that computes control updates by evaluating the expected
cost of sampled trajectories. MPPI operates in a receding-horizon fashion and is
particularly well-suited for high-dimensional, nonlinear systems such as legged
robots.

Let $x_t \in \mathbb{R}^n$ denote the system state and $u_t \in \mathbb{R}^m$ the
control input. Given a horizon length $H$ and a nominal control sequence
$\bar{u}_{0:H-1}$, MPPI generates $N$ stochastic rollouts by injecting noise into
the control sequence:
\begin{equation}
u_t^{(k)} = \bar{u}_t + \epsilon_t^{(k)}, \quad
\epsilon_t^{(k)} \sim \mathcal{N}(0, \Sigma),
\end{equation}
where $k = 1,\dots,N$ indexes the rollouts.

Each rollout is simulated forward using the system dynamics to obtain a state
trajectory $\{x_t^{(k)}\}_{t=0}^{H}$. The cumulative cost is evaluated as
\begin{equation}
S^{(k)} = \sum_{t=0}^{H-1} \ell(x_t^{(k)}, u_t^{(k)}) + \phi(x_H^{(k)}),
\end{equation}
where $\ell(\cdot)$ and $\phi(\cdot)$ denote the running and terminal cost,
respectively.

The nominal control sequence is updated using a cost-weighted average of the
noise realizations:
\begin{equation}
\bar{u}_t \leftarrow \bar{u}_t +
\sum_{k=1}^{N} w^{(k)} \epsilon_t^{(k)},
\quad
w^{(k)} =
\frac{\exp(-S^{(k)} / \lambda)}
{\sum_{j=1}^{N} \exp(-S^{(j)} / \lambda)},
\end{equation}
where $\lambda > 0$ is a temperature parameter controlling the sharpness of the
weighting.

\subsection{MPPI for Whole-Body Control of Legged Robots}

This work builds upon the MPPI-based whole-body control framework introduced in
\cite{alvarez2025real}, where MPPI is applied directly to the
high-dimensional joint-space control of legged robots. In that framework, cubic
spline parameterization is used to reduce the effective sampling dimensionality
and enforce smoothness of the control trajectories, enabling real-time execution
on hardware.

Inspired by this formulation, we adopt the same MPPI structure, cost design, and
receding-horizon execution strategy. However, instead of restricting sampling to
cubic spline parameterization, we investigate multiple structured sampling
strategies, including cubic spline, B\'ezier curve, and linear interpolation
parameterizations. The objective is to isolate and quantify the impact of the
sampling structure on MPPI performance for legged locomotion tasks.

Importantly, all methods considered in this paper share the same MPPI update rule
and differ only in how stochastic perturbations are mapped to time-continuous
control trajectories.

Algorithm~\ref{alg:mppi} summarizes the MPPI procedure used in this work.

\begin{algorithm}[t]
\caption{Model Predictive Path Integral Control}
\label{alg:mppi}
\begin{algorithmic}[1]
\REQUIRE Initial state $x_0$, initial nominal control $\bar{u}_{0:H-1}$
\FOR{each control iteration}
    \FOR{$k = 1$ to $N$}
        \STATE Sample control perturbations $\epsilon^{(k)}$
        \STATE Construct control trajectory $u^{(k)} = \bar{u} + \epsilon^{(k)}$
        \STATE Roll out system dynamics to obtain $\{x_t^{(k)}\}_{t=0}^{H}$
        \STATE Compute trajectory cost $S^{(k)}$
    \ENDFOR
    \STATE Compute importance weights $w^{(k)}$
    \STATE Update nominal control $\bar{u}_{0:H-1}$
    \STATE Execute first control input and shift horizon
\ENDFOR
\end{algorithmic}
\end{algorithm}

\section{Different Sampling Strategy in MPPI algorithm}
\subsection{Cubic Spline Sampling}
\label{sec:cubic_spline_sampling}

To generate smooth control trajectories with reduced dimensionality, we adopt a
cubic spline sampling strategy. Instead of sampling control inputs at every time
step, control values are sampled at a small number of knot points, and the full
trajectory is reconstructed using cubic spline interpolation.

Let $H$ denote the planning horizon and $m$ the action dimension. We select $K$
knot indices
\begin{equation}
\mathcal{T} = \{ t_1, t_2, \dots, t_K \},
\quad
0 = t_1 < t_2 < \dots < t_K = H-1,
\end{equation}
which are uniformly spaced over the horizon.

Given a nominal control trajectory
$\bar{u}_{0:H-1} \in \mathbb{R}^{H \times m}$,
the corresponding nominal knot values are
\begin{equation}
\bar{u}_{t_i} = \bar{u}(t_i), \quad i = 1,\dots,K.
\end{equation}

For each rollout $k$, independent Gaussian noise is applied at the knot points:
\begin{equation}
\epsilon^{(k)}_{i,j} \sim \mathcal{N}(0, \sigma_j^2),
\quad
i = 1,\dots,K,\; j = 1,\dots,m,
\end{equation}
where $\sigma_j$ denotes the noise standard deviation for the $j$-th action
dimension.

The perturbed knot values are given by
\begin{equation}
u^{(k)}_{t_i} = \bar{u}_{t_i} + \epsilon^{(k)}_{i}.
\end{equation}

For each action dimension $j$, a cubic spline is constructed by interpolating the
perturbed knot values:
\begin{equation}
u^{(k)}_j(t)
=
\mathrm{CubicSpline}\big(
\{(t_i, u^{(k)}_{t_i,j})\}_{i=1}^{K}
\big),
\quad
t = 0,\dots,H-1.
\end{equation}

Stacking all action dimensions yields the sampled control trajectory
\begin{equation}
u^{(k)}(t) =
\big[
u^{(k)}_1(t), \dots, u^{(k)}_m(t)
\big]^\top.
\end{equation}

Finally, actuator limits are enforced via element-wise clipping:
\begin{equation}
u^{(k)}(t) \leftarrow \mathrm{clip}\big(u^{(k)}(t), u_{\min}, u_{\max}\big).
\end{equation}

Cubic spline sampling enforces $C^2$ continuity of the control trajectory and
reduces the sampling dimension from $H \times m$ to $K \times m$.

\subsection{B\'ezier Curve Sampling}
\label{sec:bezier_sampling}

To generate smooth control trajectories with an explicit low-dimensional
parameterization, we employ B\'ezier curve sampling. Similar to spline-based
sampling, this approach perturbs a reduced set of control points and reconstructs
the full control trajectory through deterministic interpolation. Unlike cubic
splines, B\'ezier curves provide a global parameterization via Bernstein
polynomial bases.

Let $H$ denote the planning horizon and $m$ the action dimension. We select $K$
control point indices
\begin{equation}
\mathcal{T} = \{ t_1, t_2, \dots, t_K \},
\quad
0 = t_1 < t_2 < \dots < t_K = H-1,
\end{equation}
uniformly distributed over the horizon.

Given a nominal control trajectory
$\bar{u}_{0:H-1} \in \mathbb{R}^{H \times m}$,
the nominal control points are defined as
\begin{equation}
\bar{p}_i = \bar{u}(t_i),
\quad
i = 1,\dots,K.
\end{equation}

For each rollout $k$, Gaussian noise is applied independently to each control
point:
\begin{equation}
\epsilon^{(k)}_{i,j} \sim \mathcal{N}(0, \sigma_j^2),
\quad
i = 1,\dots,K,\; j = 1,\dots,m.
\end{equation}

The perturbed control points are given by
\begin{equation}
p^{(k)}_i = \bar{p}_i + \epsilon^{(k)}_i.
\end{equation}

The full control trajectory is reconstructed using Bernstein polynomial basis
functions. Let $n = K - 1$ denote the degree of the B\'ezier curve and
$\tau \in [0,1]$ the normalized time variable. The Bernstein basis is defined as
\begin{equation}
B_i^{n}(\tau) =
\binom{n}{i} \tau^{\,i} (1 - \tau)^{n - i},
\quad
i = 0,\dots,n.
\end{equation}

The sampled control trajectory for rollout $k$ is given by
\begin{equation}
u^{(k)}_j(t)
=
\sum_{i=0}^{n}
p^{(k)}_{i,j} \, B_i^{n}\!\left( \tfrac{t}{H-1} \right),
\quad
t = 0,\dots,H-1,
\end{equation}
for each action dimension $j$.

Stacking all action dimensions yields
\begin{equation}
u^{(k)}(t)
=
\big[
u^{(k)}_1(t), \dots, u^{(k)}_m(t)
\big]^\top.
\end{equation}

Finally, actuator limits are enforced via clipping:
\begin{equation}
u^{(k)}(t) \leftarrow \mathrm{clip}\big(u^{(k)}(t), u_{\min}, u_{\max}\big).
\end{equation}

B\'ezier curve sampling reduces the sampling dimension from $H \times m$ to
$K \times m$ while enforcing global smoothness of the control trajectory. Compared
to cubic spline sampling, B\'ezier curves offer a more interpretable
parameterization through explicit control points and guarantee that the
trajectory lies within the convex hull of these points.

\subsection{Linear Interpolation--Based Sampling}

In addition to spline-based sampling, we consider a linear interpolation--based sampling strategy as a low-complexity baseline within the MPPI framework. The objective is to reduce the dimensionality of the sampling space while preserving basic temporal structure in the control trajectories, without enforcing higher-order smoothness constraints.

Let the control horizon be $T$ and the action dimension be $d_u$. Instead of directly sampling the full control sequence $\mathbf{u}_{0:T-1} \in \mathbb{R}^{T \times d_u}$, we define a set of $K$ waypoints indexed by
\begin{equation}
\mathcal{I} = \{ i_1, i_2, \dots, i_K \}, \quad i_k \in \{0, \dots, T-1\},
\end{equation}
where the indices are uniformly distributed over the horizon:
\begin{equation}
i_k = \left\lfloor \frac{(k-1)(T-1)}{K-1} \right\rceil, \quad k = 1,\dots,K.
\end{equation}

The mean waypoints are extracted from the current mean control trajectory $\bar{\mathbf{u}}$ as
\begin{equation}
\bar{\mathbf{w}}_k = \bar{\mathbf{u}}_{i_k}, \quad k = 1,\dots,K.
\end{equation}

For each rollout $n \in \{1,\dots,N\}$, waypoint perturbations are sampled independently according to
\begin{equation}
\mathbf{w}_k^{(n)} = \bar{\mathbf{w}}_k + \boldsymbol{\epsilon}_k^{(n)}, 
\quad \boldsymbol{\epsilon}_k^{(n)} \sim \mathcal{N}(\mathbf{0}, \Sigma),
\end{equation}
where $\Sigma = \sigma^2 I$ denotes an isotropic covariance.

The full control trajectory is reconstructed via piecewise linear interpolation between adjacent waypoints:
\begin{equation}
\mathbf{u}_t^{(n)} = \mathrm{Interp}\!\left(t;\, \{(i_k, \mathbf{w}_k^{(n)})\}_{k=1}^K \right),
\quad t = 0,\dots,T-1,
\end{equation}
where the interpolation is applied independently along each action dimension. Action bounds are enforced through element-wise clipping:
\begin{equation}
\mathbf{u}_t^{(n)} \leftarrow \mathrm{clip}(\mathbf{u}_t^{(n)}, \mathbf{u}_{\min}, \mathbf{u}_{\max}).
\end{equation}

This approach reduces the stochastic sampling dimension from $T \times d_u$ to $K \times d_u$. However, due to the absence of curvature continuity, the resulting trajectories are less smooth than those produced by spline-based parameterizations.

\section{Simulation Results: Sampling Strategy Comparison}

This section presents a comparative evaluation of different control sampling parameterizations within the MPPI framework. All experiments are conducted in simulation using a Unitree-class quadruped robot model and are designed to isolate the effect of the sampling strategy on locomotion performance.

\subsection{Experimental Setup}

The MPPI controller described in the previous section is used without modification across all experiments. In particular, the system dynamics model, cost function design, temperature parameter, planning horizon, number of rollouts, and control frequency are kept identical for all methods. The only variation between experiments lies in how control perturbations are parameterized and reconstructed over the prediction horizon.

Each control strategy is evaluated on three representative locomotion tasks of increasing difficulty:
\begin{enumerate}
    \item \textbf{Flat-ground walking}: straight-line walking on level terrain toward a goal position.
    \item \textbf{Stair climbing}: ascending a staircase with discrete height changes, requiring sustained vertical motion and accurate foot placement.
    \item \textbf{Large obstacle traversal}: stepping over a tall box that induces significant kinematic and dynamic coupling between the robot body and legs.
\end{enumerate}

For each task and sampling strategy, five independent trials are conducted from identical initial conditions. A trial is considered successful if the robot reaches the target position without falling or violating joint or contact constraints.

We present three performance metrics:\begin{itemize}
    \item \emph{Success rate}: the percentage of trials that successfully reach the goal.
    \item \emph{Steps to goal}: the number of control steps required to reach the target, averaged over successful trials.
    \item \emph{Computation time}: the average wall-clock time per MPPI iteration, measured on the same hardware.
\end{itemize}

These metrics jointly characterize robustness, control efficiency, and computational cost, which are all critical for real-time whole-body control.

\subsection{Flat-Ground Walking}
\label{subsec:flat_walking}

We first evaluate the proposed sampling parameterizations on a flat-ground walking task, where the quadruped is commanded to walk straight for a distance of 1.0~m. This scenario serves as a baseline benchmark to assess gait stability, convergence speed, and computational efficiency in the absence of environmental disturbances.

Table~\ref{tab:walk_results} summarizes the quantitative results over five trials per method. The baseline MPPI with independent Gaussian sampling (\textit{Normal}) achieves only a 60\% success rate and exhibits the slowest convergence, requiring $1358.0 \pm 571.8$ steps to reach the goal. The large variance in steps reflects unstable rollouts and oscillatory joint commands, which frequently prevent reliable forward progression.

In contrast, structured sampling methods that enforce temporal coherence substantially improve both robustness and convergence. Among them, \textit{CubicSpline-k4} achieves a 100\% success rate and converges significantly faster than all other methods, reaching the goal in $444.0 \pm 30.2$ steps. This result indicates that a low-dimensional spline parameterization effectively balances expressiveness and regularization, yielding smooth and stable control sequences with minimal variance across trials.

Increasing the spline dimensionality, however, degrades robustness. While \textit{CubicSpline-k8} still improves upon the baseline in terms of convergence speed, its success rate drops to 80\% and the variance in steps increases substantially ($687.0 \pm 661.0$), suggesting sensitivity to over-parameterization under limited sampling budgets. This trend highlights the importance of controlling the effective dimensionality of the action space within MPPI.

\begin{figure}[ht]
    \centering
    \includegraphics[width=\linewidth]{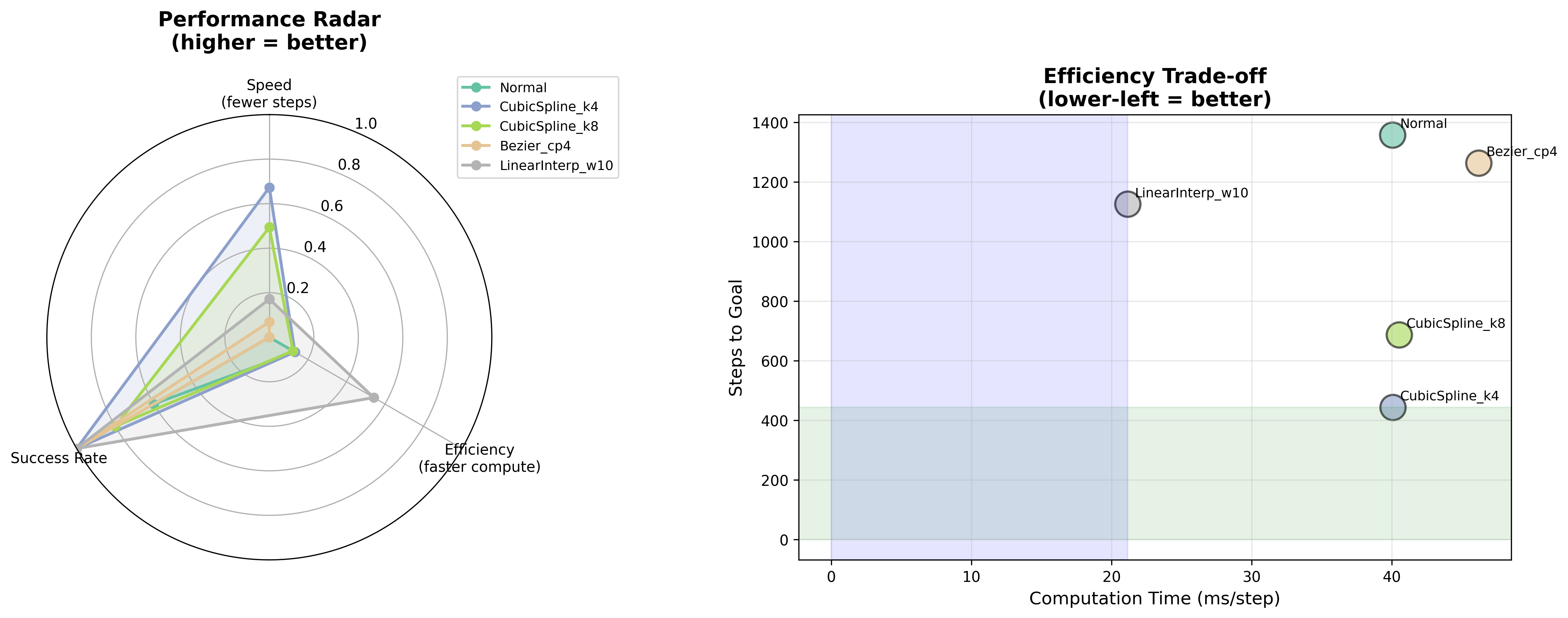}
    \caption{Performance radar and efficiency trade-off for flat-ground walking.}
    \label{fig:walk_radar}
\end{figure}

\begin{figure}[ht]
    \centering
    \includegraphics[width=\linewidth]{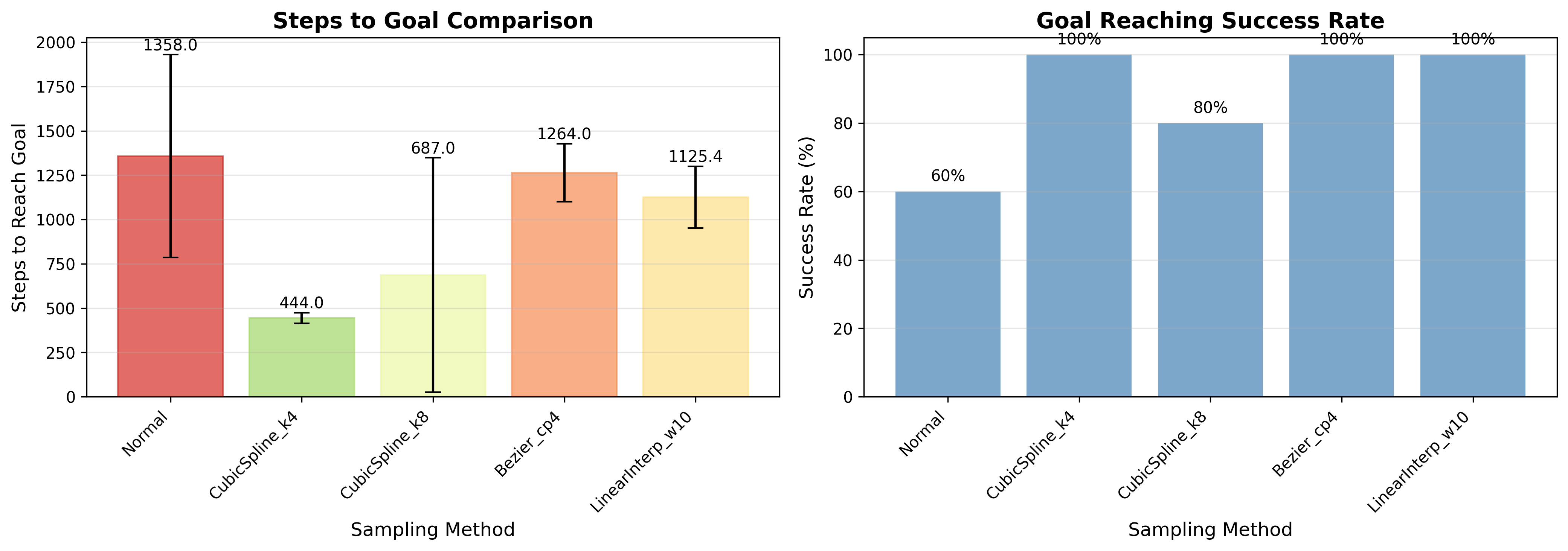}
    \caption{Steps-to-goal and success-rate comparison for different sampling methods on flat-ground walking.}
    \label{fig:walk_efficiency}
\end{figure}

Alternative structured representations exhibit different trade-offs. \textit{Bezier-cp4} and \textit{LinearInterp-w10} both achieve 100\% success rates but require over 1100 steps on average to reach the goal. Notably, \textit{LinearInterp-w10} is the most computationally efficient method, with an average computation time of $21.14 \pm 4.17$~ms per step, approximately half that of spline-based approaches, albeit at the cost of slower convergence.

These trade-offs are further illustrated in Fig.~\ref{fig:walk_efficiency}, which visualizes the relationship between computation time and convergence speed, and Fig.~\ref{fig:walk_radar}, which provides a normalized performance radar comparison across speed, efficiency, and success rate. Overall, the results demonstrate that moderate-dimensional spline parameterizations, particularly \textit{CubicSpline-k4}, offer the most favorable balance between convergence speed, robustness, and computational cost for flat-ground walking.

\begin{table}[ht]
\centering
\caption{Performance comparison on flat-ground walking.}
\label{tab:walk_results}
\begin{tabular}{lccc}
\toprule
Method & Success (\%) & Steps to Goal & Time (ms) \\
\midrule
Normal & 60 & 1358.0 $\pm$ 571.8 & 40.05 $\pm$ 13.23 \\
CubicSpline-k4 & 100 & 444.0 $\pm$ 30.2 & 40.07 $\pm$ 8.70 \\
CubicSpline-k8 & 80 & 687.0 $\pm$ 661.0 & 40.52 $\pm$ 11.21 \\
Bezier-cp4 & 100 & 1264.0 $\pm$ 163.8 & 46.19 $\pm$ 13.54 \\
LinearInterp-w10 & 100 & 1125.4 $\pm$ 174.7 & 21.14 $\pm$ 4.17 \\
\bottomrule
\end{tabular}
\end{table}

\subsection{Stair Climbing}
\label{subsec:stairs}

We next evaluate the sampling strategies on a stair-climbing task, where the quadruped ascends a staircase over a forward distance of 3.3~m. Compared to flat-ground walking, this scenario imposes substantially higher demands on long-horizon coordination, precise foothold placement, and vertical body motion, making it a stringent test of temporal consistency in the sampled control sequences.

Table~\ref{tab:stairs_results} reports the quantitative performance over five trials per method. The baseline MPPI with independent Gaussian sampling (\textit{Normal}) fails entirely, achieving a 0\% success rate and consistently reaching the maximum episode length of 4000 steps. Despite its relatively low per-step computation time, the lack of temporal structure leads to incoherent action sequences that are unable to sustain stable climbing behavior. A similar failure mode is observed for \textit{Bezier-cp4}, which also achieves a 0\% success rate while incurring the highest computational cost among all methods.

In contrast, spline-based sampling demonstrates a decisive advantage in this long-horizon task. Both \textit{CubicSpline-k4} and \textit{CubicSpline-k8} achieve 100\% success rates, reliably completing the stair ascent in approximately 1000 steps on average. Among them, \textit{CubicSpline-k4} converges slightly faster, with $1010.6 \pm 232.8$ steps compared to $1031.6 \pm 332.4$ for \textit{CubicSpline-k8}, while also exhibiting lower variance. This result again highlights that moderate-dimensional spline parameterizations provide sufficient expressiveness for complex maneuvers without introducing unnecessary sensitivity to sampling noise.

The \textit{LinearInterp-w10} method achieves partial success, with a 60\% success rate, but requires significantly more steps to reach the goal ($2817.2 \pm 1032.3$) and exhibits large trial-to-trial variability. While its lower computation time makes it attractive from an efficiency standpoint, the piecewise-linear structure appears insufficient to capture the smooth and coordinated motion required for reliable stair climbing.

\begin{figure}[t]
    \centering
    \includegraphics[width=\linewidth]{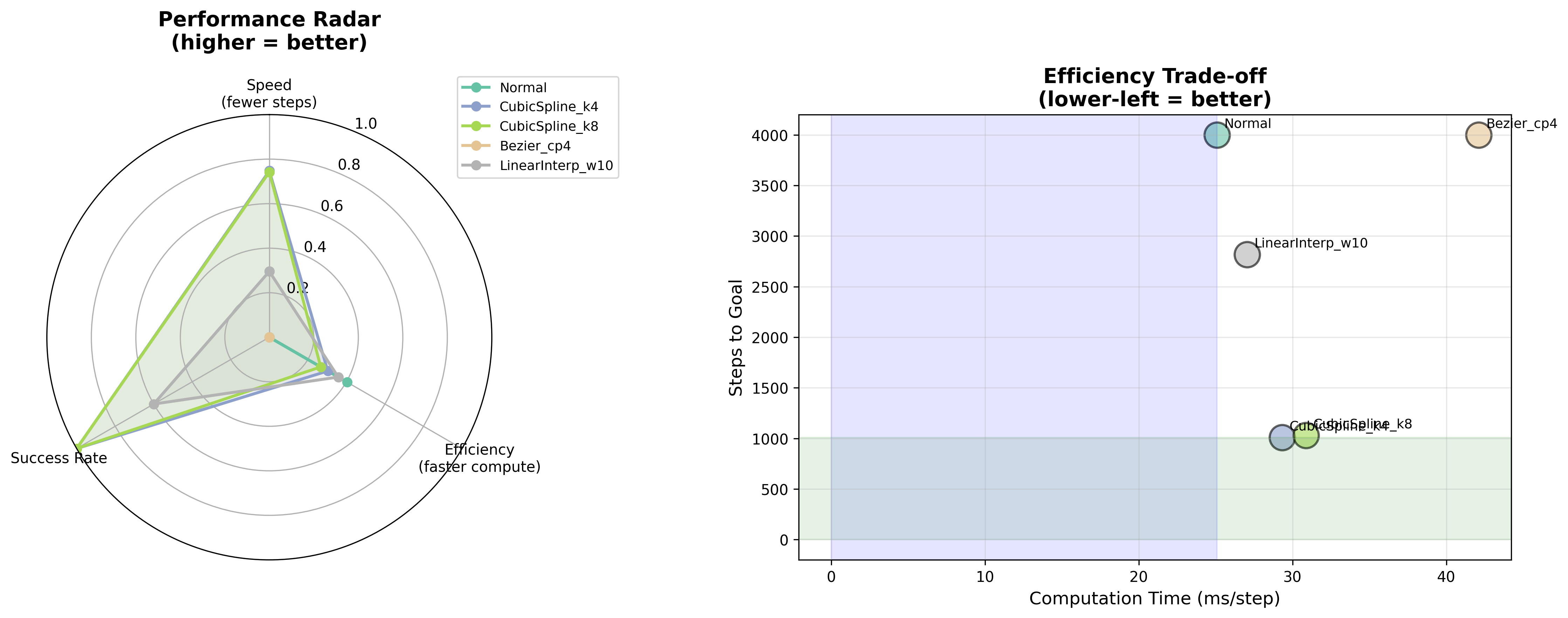}
    \caption{Performance radar and efficiency trade-off for the stair-climbing task.}
    \label{fig:stairs_radar}
\end{figure}

\begin{figure}[t]
    \centering
    \includegraphics[width=\linewidth]{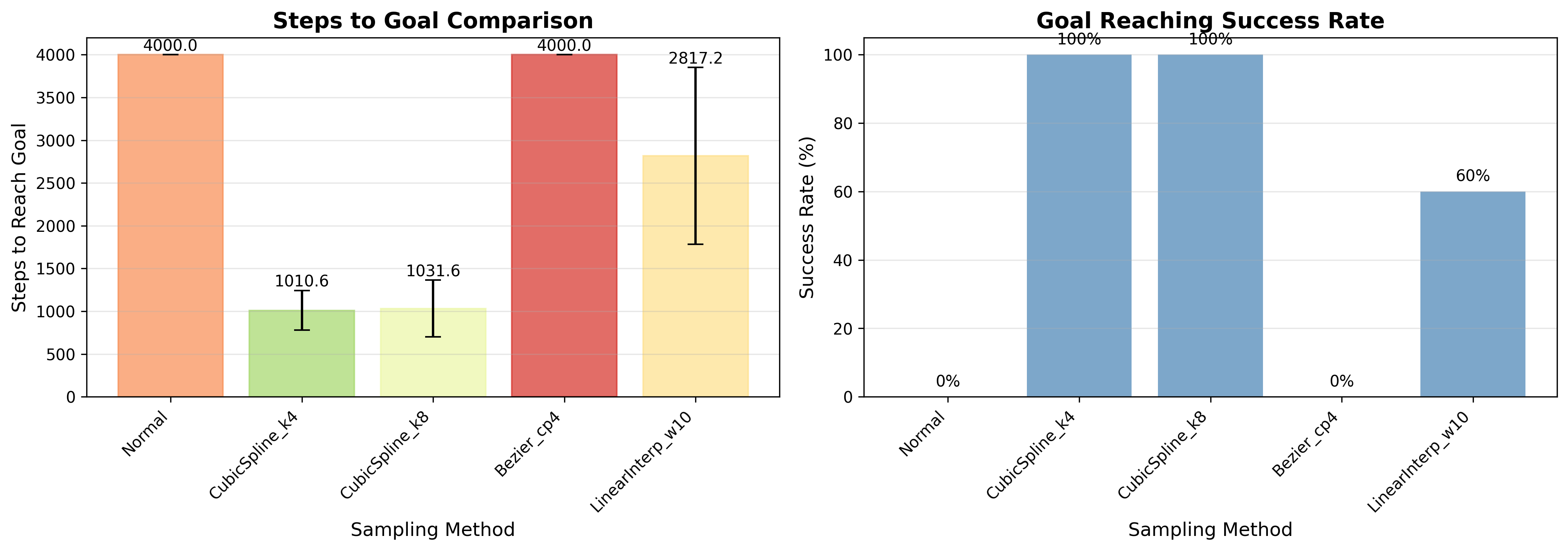}
    \caption{Steps-to-goal and success-rate comparison for different sampling methods on the stair-climbing task.}
    \label{fig:stairs_efficiency}
\end{figure}

Figure~\ref{fig:stairs_efficiency} illustrates the trade-off between computation time and convergence speed, clearly separating successful spline-based methods from failure cases. Figure~\ref{fig:stairs_radar} further summarizes the normalized performance across convergence speed, efficiency, and success rate. Overall, these results demonstrate that temporally smooth, low-dimensional spline parameterizations are not merely beneficial but essential for solving contact-rich, long-horizon locomotion tasks such as stair climbing within the MPPI framework.

\begin{table}[t]
\centering
\caption{Performance comparison on stair climbing.}
\label{tab:stairs_results}
\begin{tabular}{lccc}
\toprule
Method & Success (\%) & Steps to Goal & Time (ms) \\
\midrule
Normal & 0 & 4000.0 $\pm$ 0.0 & 25.08 $\pm$ 5.79 \\
CubicSpline-k4 & 100 & 1010.6 $\pm$ 232.8 & 29.34 $\pm$ 6.89 \\
CubicSpline-k8 & 100 & 1031.6 $\pm$ 332.4 & 30.87 $\pm$ 11.61 \\
Bezier-cp4 & 0 & 4000.0 $\pm$ 0.0 & 42.11 $\pm$ 11.94 \\
LinearInterp-w10 & 60 & 2817.2 $\pm$ 1032.3 & 27.04 $\pm$ 8.52 \\
\bottomrule
\end{tabular}
\end{table}

\subsection{Big Box Obstacle Traversal}
\label{subsec:big_box}

We finally evaluate the sampling strategies on a large obstacle traversal task, where the quadruped must step over a tall box while advancing 1.0~m forward. Unlike stair climbing, this scenario emphasizes short-horizon, highly coupled whole-body coordination, requiring rapid adaptation of body posture and foot placement within a limited spatial window.

Table~\ref{tab:box_results} summarizes the performance over five trials per method. The baseline MPPI with independent Gaussian sampling (\textit{Normal}) again fails in all trials, achieving a 0\% success rate and consistently reaching the maximum episode length of 4000 steps. Despite a moderate computation time, the lack of structured temporal coupling prevents the controller from producing coordinated lifting and landing motions necessary to clear the obstacle.

Among all methods, \textit{CubicSpline-k4} achieves the best overall performance, with a 100\% success rate and the fewest steps to reach the goal ($955.4 \pm 739.8$). Although the variance is higher than in flat-ground walking, this is expected due to the inherently discontinuous contact events during obstacle traversal. Importantly, \textit{CubicSpline-k4} remains both robust and computationally efficient, with an average computation time of $23.25 \pm 5.01$~ms per step.

Increasing the spline dimensionality leads to reduced robustness in this task. \textit{CubicSpline-k8} achieves an 80\% success rate and requires more steps on average ($1332.4 \pm 1379.3$), despite being the fastest method in terms of per-step computation time. This result suggests that while higher-dimensional parameterizations may offer additional flexibility, they can also introduce unnecessary variability in short-horizon, contact-rich maneuvers.

The remaining structured methods exhibit intermediate performance. \textit{LinearInterp-w10} reaches a 60\% success rate with large variance in steps ($2239.4 \pm 1461.0$), indicating inconsistent obstacle negotiation. \textit{Bezier-cp4} performs the worst among structured approaches, succeeding in only 40\% of trials and often failing to generate sufficiently aggressive or well-timed motions to clear the obstacle.

\begin{figure}[ht]
    \centering
    \includegraphics[width=\linewidth]{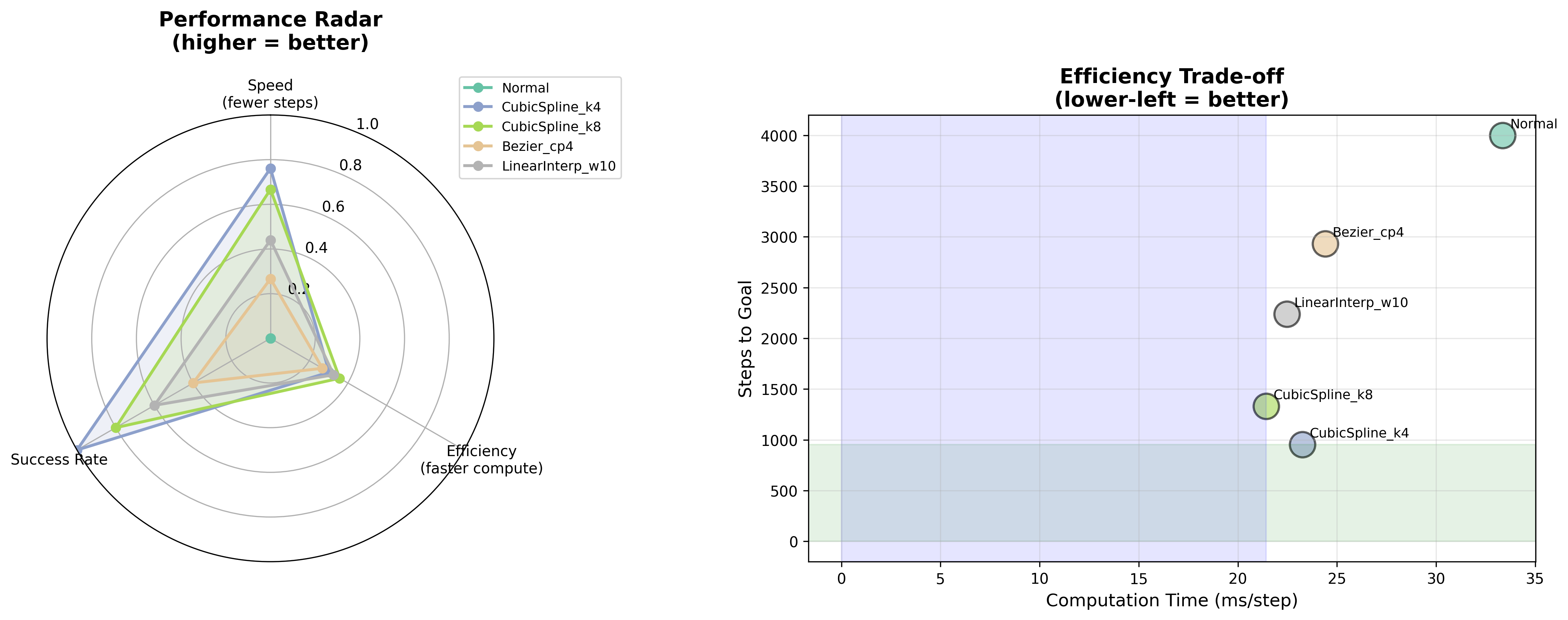}
    \caption{Performance radar and efficiency trade-off for the big box obstacle traversal task.}
    \label{fig:bigbox_radar}
\end{figure}

\begin{figure}[ht]
    \centering
    \includegraphics[width=\linewidth]{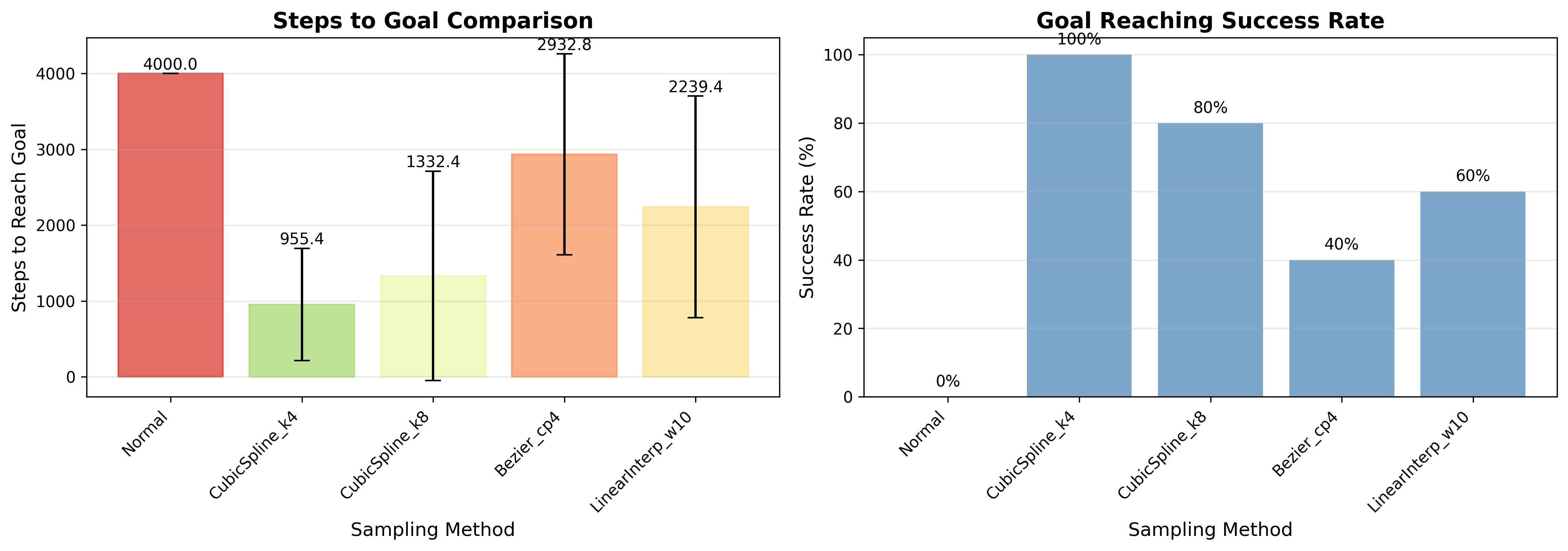}
    \caption{Steps-to-goal and success-rate comparison for different sampling methods on the stair-climbing task.}
    \label{fig:bigbox_efficiency}
\end{figure}

Figure~\ref{fig:bigbox_efficiency} illustrates the trade-off between convergence speed and computational cost, clearly separating successful spline-based methods from failure cases. Figure~\ref{fig:bigbox_radar} further provides a normalized comparison across speed, efficiency, and success rate. Taken together, these results demonstrate that for short-horizon but highly coordinated behaviors, such as large obstacle traversal, low-dimensional spline parameterizations again provide the most favorable balance between expressiveness, robustness, and sample efficiency within MPPI.

\begin{table}[t]
\centering
\caption{Performance comparison on big box traversal.}
\label{tab:box_results}
\begin{tabular}{lccc}
\toprule
Method & Success (\%) & Steps to Goal & Time (ms) \\
\midrule
Normal & 0 & 4000.0 $\pm$ 0.0 & 33.36 $\pm$ 8.79 \\
CubicSpline-k4 & 100 & 955.4 $\pm$ 739.8 & 23.25 $\pm$ 5.01 \\
CubicSpline-k8 & 80 & 1332.4 $\pm$ 1379.3 & 21.43 $\pm$ 3.89 \\
Bezier-cp4 & 40 & 2932.8 $\pm$ 1323.5 & 24.40 $\pm$ 4.39 \\
LinearInterp-w10 & 60 & 2239.4 $\pm$ 1461.0 & 22.47 $\pm$ 4.58 \\
\bottomrule
\end{tabular}
\end{table}

\subsection{Simulation Summary}
In summary, the simulation results demonstrate that the choice of sampling parameterization within the MPPI framework has a pronounced impact on both task performance and robustness across diverse locomotion scenarios. While all evaluated strategies are capable of generating feasible motions on flat terrain, clear differences emerge in more challenging environments such as stair climbing and large obstacle traversal. Parameterizations that better capture the underlying structure of the control sequence consistently achieve lower task costs, improved tracking accuracy, and reduced control variability, particularly under contact-rich and highly nonlinear dynamics. Moreover, these strategies exhibit superior sample efficiency, enabling more stable behaviors without increasing the computational budget. Collectively, these findings indicate that informed sampling design is a critical factor for scaling MPPI to complex legged locomotion tasks, and they motivate the use of structured or low-dimensional sampling spaces when deploying MPPI on real quadruped platforms.

\section{Conclusion and Future Work}

This paper investigates the role of sampling strategy design within the MPPI framework for legged robot locomotion. Through extensive simulations on a quadruped robot, we tested and compared MPPI controllers equipped with different control sampling strategies across multiple terrains and task scenarios. The results demonstrate that, although MPPI provides a unified stochastic optimal control formulation, its practical performance on legged robots is strongly influenced by the choice of sampling parameterization. In particular, structured sampling strategies lead to improved task performance, enhanced robustness in contact-rich environments, and increased sample efficiency compared to unstructured sampling approaches. These findings highlight sampling design as a critical component for effectively applying MPPI to complex legged locomotion problems.

Building on the insights obtained in this study, future work will focus on a broader investigation of sampling strategies for optimal control. We aim to analyze how different sampling parameterizations affect the exploration–exploitation trade-off, convergence properties, and computational efficiency of stochastic optimal control algorithms, including but not limited to MPPI. In addition, we plan to explore adaptive and learning-based sampling mechanisms that can adjust to task requirements and system dynamics online. Ultimately, this line of research seeks to establish principled guidelines for sampling design that enable scalable and reliable deployment of sampling-based optimal control methods on real-world robotic systems.

\bibliographystyle{ieeetr}
\bibliography{multi}

@inproceedings{williams2017information,
  title={Information theoretic MPC for model-based reinforcement learning},
  author={Williams, Grady and Wagener, Nolan and Goldfain, Brian and Drews, Paul and Rehg, James M and Boots, Byron and Theodorou, Evangelos A},
  booktitle={2017 IEEE international conference on robotics and automation (ICRA)},
  pages={1714--1721},
  year={2017},
  organization={IEEE}
}

@article{williams2018information,
  title={Information-theoretic model predictive control: Theory and applications to autonomous driving},
  author={Williams, Grady and Drews, Paul and Goldfain, Brian and Rehg, James M and Theodorou, Evangelos A},
  journal={IEEE Transactions on Robotics},
  volume={34},
  number={6},
  pages={1603--1622},
  year={2018},
  publisher={IEEE}
}

@article{williams2015model,
  title={Model predictive path integral control using covariance variable importance sampling},
  author={Williams, Grady and Aldrich, Andrew and Theodorou, Evangelos},
  journal={arXiv preprint arXiv:1509.01149},
  year={2015}
}

@inproceedings{williams2016aggressive,
  title={Aggressive driving with model predictive path integral control},
  author={Williams, Grady and Drews, Paul and Goldfain, Brian and Rehg, James M and Theodorou, Evangelos A},
  booktitle={2016 IEEE international conference on robotics and automation (ICRA)},
  pages={1433--1440},
  year={2016},
  organization={IEEE}
}

@inproceedings{hutter2016anymal,
  title={Anymal-a highly mobile and dynamic quadrupedal robot},
  author={Hutter, Marco and Gehring, Christian and Jud, Dominic and Lauber, Andreas and Bellicoso, C Dario and Tsounis, Vassilios and Hwangbo, Jemin and Bodie, Karen and Fankhauser, Peter and Bloesch, Michael and others},
  booktitle={2016 IEEE/RSJ international conference on intelligent robots and systems (IROS)},
  pages={38--44},
  year={2016},
  organization={IEEE}
}

@inproceedings{alvarez2025real,
  title={Real-time whole-body control of legged robots with model-predictive path integral control},
  author={Alvarez-Padilla, Juan and Zhang, John Z and Kwok, Sofia and Dolan, John M and Manchester, Zachary},
  booktitle={2025 IEEE International Conference on Robotics and Automation (ICRA)},
  pages={14721--14727},
  year={2025},
  organization={IEEE}
}

@inproceedings{tao2022path,
  title={Path integral methods with stochastic control barrier functions},
  author={Tao, Chuyuan and Yoon, Hyung-Jin and Kim, Hunmin and Hovakimyan, Naira and Voulgaris, Petros},
  booktitle={2022 IEEE 61st Conference on Decision and Control (CDC)},
  pages={1654--1659},
  year={2022},
  organization={IEEE}
}

@article{tao2025guided,
  title={Guided Sampling-Based Motion Planning Algorithm for Dynamic and Obstacle Cluttered Environments},
  author={Tao, Chuyuan and Kim, Hunmin and Yoon, Hyung-Jin and Voulgaris, Petros and Hovakimyan, Naira},
  journal={Journal of Guidance, Control, and Dynamics},
  pages={1--18},
  year={2025},
  publisher={American Institute of Aeronautics and Astronautics}
}

@article{wang2025dbas,
  title={DBaS-Log-MPPI: Efficient and Safe Trajectory Optimization via Barrier States},
  author={Wang, Fanxin and Jiang, Haolong and Tao, Chuyuan and Wan, Wenbin and Cheng, Yikun},
  journal={arXiv preprint arXiv:2504.06437},
  year={2025}
}

@article{cheng2022improving,
  title={Improving the Robustness of Reinforcement Learning Policies With $\mathcal{L}_1$ Adaptive Control},
  author={Cheng, Yikun and Zhao, Pan and Wang, Fanxin and Block, Daniel J and Hovakimyan, Naira},
  journal={IEEE Robotics and Automation Letters},
  volume={7},
  number={3},
  pages={6574--6581},
  year={2022},
  publisher={IEEE}
}

@inproceedings{cheng2023safe,
  title={Safe and efficient reinforcement learning using disturbance-observer-based control barrier functions},
  author={Cheng, Yikun and Zhao, Pan and Hovakimyan, Naira},
  booktitle={Learning for Dynamics and Control Conference},
  pages={104--115},
  year={2023},
  organization={PMLR}
}

@article{theodorou2010generalized,
  title={A generalized path integral control approach to reinforcement learning},
  author={Theodorou, Evangelos and Buchli, Jonas and Schaal, Stefan},
  journal={The Journal of Machine Learning Research},
  volume={11},
  pages={3137--3181},
  year={2010},
  publisher={JMLR. org}
}

@article{todorov2009efficient,
  title={Efficient computation of optimal actions},
  author={Todorov, Emanuel},
  journal={Proceedings of the national academy of sciences},
  volume={106},
  number={28},
  pages={11478--11483},
  year={2009},
  publisher={National Academy of Sciences}
}

@article{lowrey2018plan,
  title={Plan online, learn offline: Efficient learning and exploration via model-based control},
  author={Lowrey, Kendall and Rajeswaran, Aravind and Kakade, Sham and Todorov, Emanuel and Mordatch, Igor},
  journal={arXiv preprint arXiv:1811.01848},
  year={2018}
}

@article{testouri2023towards,
  title={Towards a safe real-time motion planning framework for autonomous driving systems: An mppi approach},
  author={Testouri, Mehdi and Elghazaly, Gamal and Frank, Raphael},
  journal={arXiv preprint arXiv:2308.01654},
  year={2023}
}

@book{betts2010practical,
  title={Practical methods for optimal control and estimation using nonlinear programming},
  author={Betts, John T},
  year={2010},
  publisher={SIAM}
}

@inproceedings{tassa2014control,
  title={Control-limited differential dynamic programming},
  author={Tassa, Yuval and Mansard, Nicolas and Todorov, Emo},
  booktitle={2014 IEEE International Conference on Robotics and Automation (ICRA)},
  pages={1168--1175},
  year={2014},
  organization={IEEE}
}

@misc{rathod2021terrain,
  title={Terrain Aware Model Predictive Control for Legged Locomotion},
  author={Rathod, Niraj and Bratta, Angelo and Focchi, Michele and Zanon, Mario and Villarreal, Octavio and Semini, Claudio and Bemporad, Alberto},
  year={2021},
  publisher={ArXiv}
}

@article{chatzinikolaidis2020contact,
  title={Contact-implicit trajectory optimization using an analytically solvable contact model for locomotion on variable ground},
  author={Chatzinikolaidis, Iordanis and You, Yangwei and Li, Zhibin},
  journal={IEEE Robotics and Automation Letters},
  volume={5},
  number={4},
  pages={6357--6364},
  year={2020},
  publisher={IEEE}
}

@article{gangapurwala2022rloc,
  title={Rloc: Terrain-aware legged locomotion using reinforcement learning and optimal control},
  author={Gangapurwala, Siddhant and Geisert, Mathieu and Orsolino, Romeo and Fallon, Maurice and Havoutis, Ioannis},
  journal={IEEE Transactions on Robotics},
  volume={38},
  number={5},
  pages={2908--2927},
  year={2022},
  publisher={IEEE}
}
\end{document}